# GEOMETRY OF TIME AND DIMENSIONALITY OF SPACE

M. SANIGA
*International Solvay Institutes for Physics and Chemistry, ULB,*
*Campus Plaine, CP–231, Blvd du Triomphe, 1050 Brussels, Belgium*
*&*
*Astronomical Institute of the Slovak Academy of Sciences,*
*05960 Tatranská Lomnica, Slovak Republic*

## 1. Introduction

It goes without saying that a profound mystery lies behind the conventional notions of space and time. Thus, for example, the fact that there are *three* macroscopic dimensions of space was rigorously proved as early as the great Ptolemy some thousand years ago, yet contemporary science is still lacking any deeper and theoretically well-founded insight into the origin of this puzzling number. Deeper than the enigma of (the dimensionality of) space seems to be that of (the nature of) time. Here, there even exists a sharp contradiction between the way we perceive time and what modern physical theories tell us about the concept. To our senses, time appears to "flow," "pass," proceed inexorably from the past through (the unique moment of) the present into the future – the fact commonly known as the *arrow* of time. Yet, almost all the fundamental equations of physics are strictly time-reversible and, in addition, they do not leave any proper place for the concept of the present, the "now." This failure of current physical theories to properly account for the observed macroscopic dimensionality of space and the intricate nature of time is, in our opinion, asking for a serious revision of the generally adopted physical paradigms about the concepts in question. We think that there is a strong need for the representation of space and time that is more adherent to our perceptions and which includes, in particular, the *ir*reversibility of change.

Modern theoretical physics has entered the territory of scientific inquiry that lies so far from ordinary experience that there exists no rigorous observational/ experimental guide to be followed. The only means physicists have at hand to navigate through this region is mere appeal of abstract and often counter-intuitive mathematical principles. Yet, sticking to mathematical beauty alone may not necessarily be a proper path leading to a discovery of new, more fundamental physical laws. For the history of science, and physics in particular, teaches us a very important lesson that novel, revolutionary ideas and paradigm shifts were almost always preceded and accompanied by new evidence from observations and experiments that had accumulated over particular periods. So why not to listen to this lesson again? This is precisely the strategy we adopted some fifteen years ago, soon after we became familiar with a fascinating and extremely thought-provoking topic of psychopathology of space and time – the latter being a generic term for all "peculiar," or "abnormal," perceptions of space and time as invariably reported by people suffering serious mental psychoses as well as by other subjects experiencing so-called "altered" states of consciousness [1–14]. Over the years, a vast body of literature has accumulated on the topic (see references in [21–23]) so that there can already be seen a definite pattern in the qualitative structure of these pathological space-time constructs. Already in the normal state of health there are, every now and then, aberrations of subjective time such as acceleration or deceleration of the lapse of time. Under severe mental disturbances these anomalies/peculiarities become more pronounced. The flux of time may even cease completely (the sensations usually described as "time standing still," or "suspended," "arrested" time), or expand without limit (the feelings of "everlasting now," "eternity"). In some cases, time's flow may be experienced as discontinuous, fragmented or even reversing its direction. Finally, in most extreme cases, time as a dimension is transcended, or simply non-existent ("atemporal," "timeless" states). The sense of space is likewise powerfully affected. Space can appear "amplified" or "compressed," "condensed" or "rarefied," or even changing its dimensionality; it can, for example, become just two dimensional ("flat"), acquire another dimensions, or be simply reduced to a dimensionless point in consciousness.

Obviously, it would be an utterly hopeless task if we tried to explain these fascinating space-time constructs in terms of physics. Hence, a *conceptually* new framework is required to handle these phenomena. Some years ago we put forward a theory that seems to be very promising in this respect [15–23]. This pencil theory was originally motivated by and aimed at a deeper insight into the puzzling discrepancy between perceptional and physical aspects of time. Yet, we soon realized that it also has an important bearing on the problem of the dimensionality of space. Namely, we found out that there seems to exist an intricate relation between our *sense* of time and the observed *number* of spatial dimensions [15,17,21,22]. Mathematically, this property is substantiated by the fact that we treat time and space from the very beginning as standing on topologically different footings. As for their "outer" appearance, both the types of dimension are identical, being regarded as pencils, i.e. linear, single-infinite aggregates of constituting elements. It is their "inner" structure where the difference comes in: the constituting element ("point") of a spatial dimension is a *line*, whereas that of the time dimension is a (proper) *conic*.

The algebraic geometrical setting of our debut theory was a projective plane. The theory acquired a qualitatively new standing when we raised the dimensionality of the setting by one, i.e. moved into a projective space, and identified the pencils in question with those of the fundamental configurations of certain Cremona transformations [24–26]. The 3+1 macroscopic dimensionality of space-time was demonstrated to uniquely follow from the structure of the so-called *quadro-cubic* Cremona transformations – the *simplest* non-trivial, non-symmetrical Cremona transformations in a projective space of three dimensions [24,25]. In addition, these transformations were also found to fix the type of the pencil of fundamental conics, i.e. the geometry of the time dimension, and to provide us with a promising conceptual basis for eventual reconciliation between two extreme views of space-time, viz. physical and psychological.

The paper gives a succinct exposition of this generalized theory. After introducing its fundamental postulates and highlighting essentials of space Cremona transformations, we review basic properties of the corresponding "Cremonian" space-times, stressing particularly those items where the departure from generally accepted views/paradigms is substantial. The presentation is rather non-technical, with the hope of being accessible to scientists of various disciplines and diverse mathematical background. The reader who wishes to go deeper into the mathematical formalism employed is referred to our papers [17,24,25].

## 2. Cremonian (Pencil-)Space-Times

Let us consider two distinct, incident lines in a projective space. These define a unique plane (they both share) and a unique point (their intersection).[1] The two lines define also a unique *pencil of lines*, i.e. the linear, single-infinite set of lines lying in the plane and passing through the point. It is our first fundamental postulate that each of the observed dimensions of space is isomorphic to a pencil of lines [15–17]. Next, let us take a plane and two distinct conics lying in it. These define a unique *pencil of conics*, i.e. the linear, single-parametrical aggregate of conics confined to the plane and passing through the points shared by the two conics (called the base points of the pencil). Our second basic postulate is that the structure of the time dimension is identical to that of a specific pencil of conics, each proper conic standing for a single event [15–17]. As any two coplanar lines have always one, and only one, point in common, there exists only one projective type of a pencil of lines. A different situation is encountered in the case of conics' pencils as two different conics situated in the same (projective) plane have four points in common, of which some (or all) may coincide, or be pair-wise imaginary: in the case where all the points are real we find as many as five projectively distinct types of pencil of conics – as illustrated in Figure 1. So, in our pencil-approach space has a simpler (less complex) structure than time – a feature conforming nicely to our sensual perception.

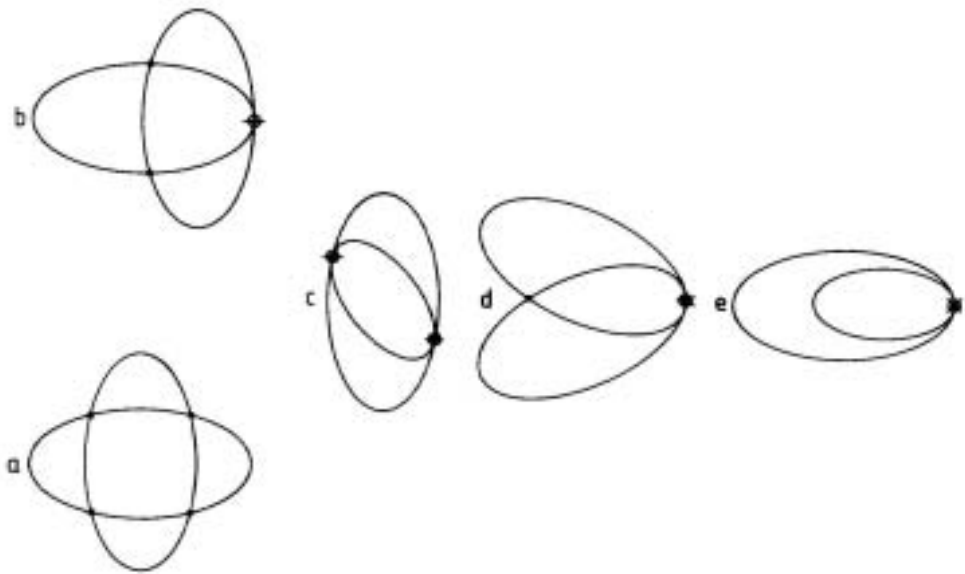

*Figure 1*. The five projectively distinct types of pencil of conics in the case where all the base points are real: *a* – all the four points distinct; *b* – two points distinct, one double; *c* – two distinct points, each of multiplicity two; *d* – one single and one triple point; *e* – one point of multiplicity four.

Clearly, any pencil of lines may serve as a potential spatial dimension and, similarly, any pencil of conics can be taken to represent the time dimension. Our pencil space-time is thus originally infinite-dimensional and lacking any definite link between time and space. We therefore need a mechanism that would, on the one side, break this symmetry down to what we really observe and, on the other, induce a unique coupling between the two kinds of dimension. It is here where Cremona transformations are invoked to do the job [24–26].

---

[1] The terminology, symbols and notation used here are the same as in [17] and [24].

A space Cremona transformation is a one-to-one (birational) correspondence between the points of two projective spaces [24,27,28]. The transformation is determined in all essentials by giving, in either space, a *homaloidal* web of surfaces, i.e. a linear, triple-infinite family of rational surfaces of which any three members have only one free (variable) intersection; these homaloidal surfaces are mapped by the transformation into the planes of the other space. The character of a homaloidal web is completely specified by the properties of its *base* configuration, that is, by the set of points that are common to every member of the web. A base point is an exceptional element of the web in the sense that it makes the equations of the corresponding Cremona transformation illusory. Its image (homologue) in the other space is thus not a single point, but a locus, either a curve or a surface, called the *fundamental* element; it is the totality of these, the fundamental configuration, whose structure turns out to be of paramount importance.

To clarify and substantiate the point just made, let us have a look at the structure of such a configuration for the simplest among asymmetrical space Cremona transformations – the transformation generated, in one of the spaces, by a homaloidal web of quadratic surfaces (quadrics) whose base manifold consists of a real line, $\mathcal{L}^B$, and three distinct, non-collinear real points $B_k$ ($k$=1,2,3), none being incident with the line in question [24,27,28]. It is an utterly amazing thing to see that this fundamental configuration comprises just *three* pencils of *lines*, viz. the ones located in the planes $B_k\mathcal{L}^B$ and centered at the points $B_k$, and just *one* pencil of *conics*, that situated in the plane $B_1B_2B_3$ and having for the base points the three points $B_k$ and the point L at which the line $\mathcal{L}^B$ meets the plane in question – as depicted in Figure 2, *left*.

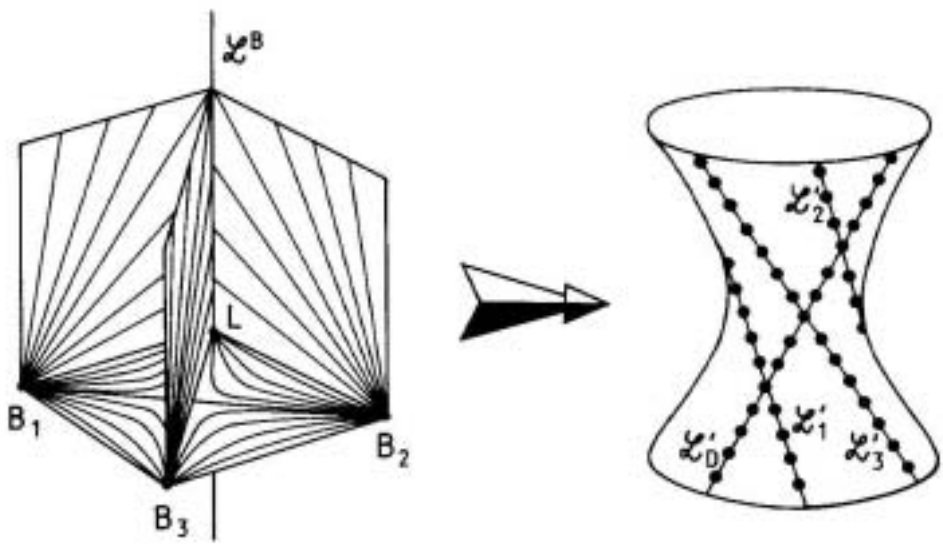

*Figure 2*. A schematic sketch of (the relation between) the structure of the fundamental configuration of the homaloidal web of quadrics featuring a real base line and three distinct real base points (*left*) and that of the base configuration of the associated web of ruled cubic surfaces (*right*). The symbols and notation are given in the text.

So, there can be nothing more natural than assuming that the space-time as perceived by our senses has the structure of the above described fundamental configuration: since once adopting this tenet, we have *a nice explanation not only why it features four macroscopic dimensions (four "fundamental" pencils), but also why three of them (spatial, generated by pencils of lines) are of a qualitatively different nature than the remaining one (time, represented by a pencil of conics)* [24]. However, it is not only the right number of macro-dimensions and their correct "ratio" that stem naturally from this picture. It is also a definite coupling between spatial dimensions and time, as envisaged above. For the vertices of the three fundamental lines' pencils ($B_k$) not only define the plane of location of the fundamental conics, but, together with the point L, they also uniquely specify the type of their pencil, i.e. the extrinsic geometry of the time dimension. As this coupling between time and space represents a considerable departure from that based on relativity theory, we shall examine it in more detail later on, when the reader is more acquainted with the approach.

The picture just outlined is obviously only one side of the coin for we cannot ignore the role played by the image of the fundamental configuration in the second ("primed") projective space, i.e. by the *base* configuration of the associated homaloidal web of surfaces. As shown in detail in [24,27,28], this inverse homaloidal web consists of surfaces of the third degree (cubics). The cubics are ruled, i.e. each contains a single-infinity of lines, and have in common four lines: three of them, $\mathcal{L}'_k$ ($k$=1,2,3), are mutually skew (disjoint), whereas the fourth one, $\mathcal{L}'_D$, is incident with each of the three – as portrayed in Figure 2, *right*.[2] Not only does the line $\mathcal{L}'_D$ stand apart from the lines $\mathcal{L}'_k$, being their common transversal, but it also differs from them in another crucial aspect: it is singular (double) for every cubic of the web, whilst the other three lines are ordinary (simple). Although not pronounced to such a degree as in the fundamental configuration, a three-to-one splitting is thus inherent also in the structure of the base system of the inverse web. That this must be so is not difficult to understand because (as also elucidated in Figure 2) the fundamental lines of the pencils in $B_k\mathcal{L}^B$ correspond to the points of the lines $\mathcal{L}'_k$, respectively, and the fundamental conics in the plane $B_1B_2B_3$ answer to the points of the remaining line, $\mathcal{L}'_D$ [24,27,28]. We must therefore regard the base configuration of the web of cubics as another viable representation of space-time, in no way less prominent than the previous one. Our daring hypothesis is that this configuration underlies qualitatively the physical conception of space-time [24]. A principal justification for such a claim goes as follows. From a general relativist's point of view, there is no distinction between time and space as far as their internal structure is concerned; the only difference between the two is embodied in the (Lorentz) signature of the metric tensor on the underlying differentiable manifold. And this is indeed very similar to what our "base" space-time exhibits, as all the four dimensions are there represented by lines; yet, the time coordinate, represented by $\mathcal{L}'_D$, has apparently a different standing than the three dimensions of space, generated by the lines $\mathcal{L}'_k$, $k$=1,2,3.

We have thus found two non-equivalent, yet robust on their own, Cremonian views of the macroscopic spatio-temporal fabric [24]. One, based on the properties of the fundamental configuration of the specific homaloidal web of quadrics (Figure 2, *left*), is characterized by a more pronounced difference between the spatial dimensions and time and, therefore, more appropriate when dealing with space-time as imprinted in our consciousness (the "subjective" view). The other, grounded in the structure of the base configuration of the associated web of cubics (Figure 2, *right*), features a less marked distinction between the spatial dimensions and time and is, so, more akin to the physical picture of space-time (the "objective" view). The two representations are intricately linked to each other, the link being mediated by the particular type of Cremona transformation, the one that sends the quadrics of the web in question into the planes of the other space. This transformation is algebraically elegant and geometrically simple [24,27,28], and may thus offer extraordinary promise for being an important initial stepping-stone towards bridging the gap between two crucial, but so far so poorly reconciled, fields of the scientific inquiry, viz. physics and psychology.

---

[2] The four lines lie on a *unique* quadric (represented in Figure 2 as a hyperboloid of one sheet).

After demonstrating that the correct macroscopic dimensionality (4) and signature (3+1) of the Universe are both crucial characteristics of our Cremonian space-times, we now proceed, as promised, to have a closer look at how the three spatial pencil-dimensions are coupled to the time pencil-coordinate. The best way to illustrate this point is to return to the generic homaloidal web of quadrics and see what happens if, for example, one of the isolated base points, say $B_3$, approaches the base line $\mathcal{L}^B$ until the two get ultimately incident. As shown in detail in [25], the resulting fundamental configuration still comprises three distinct pencils of lines and a single pencil of conics: however, one of the pencils of lines now incorporates the base line $\mathcal{L}^B$ (that centered at the point $B_3$, and henceforth called "extraordinary"), and stands thus slightly apart from the other two ("ordinary"), which do not – see Figure 3, *left*. In the other space, this asymmetry answers to the fact that two of the three simple base lines ($\mathcal{L}'_1$ and $\mathcal{L}'_2$) meet each other, while the third one ($\mathcal{L}'_3$) is skew with either – see Figure 3, *right*.

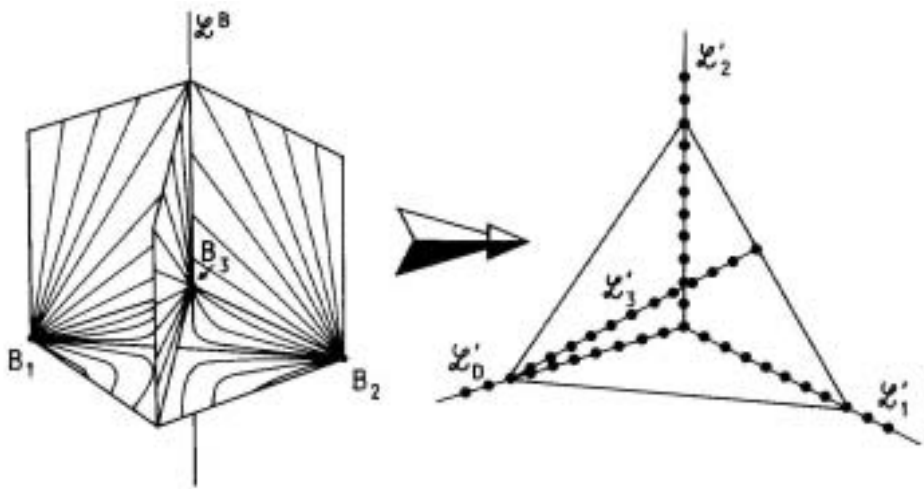

*Figure 3*. An illustrative sketch of the fundamental configuration of a homaloidal web of quadrics sharing a real base line and three isolated base points of which one ($B_3$) falls on the line in question (*left*) and the base configuration of the inverse homaloidal web of cubics (*right*). The symbols are identical to those of the previous figure.

The corresponding Cremonian space-times thus both exhibit an intriguing "anisotropy" among the spatial dimensions, where one dimension has a slightly different footing than the other two. It is, however, the "fundamental" ("subjective") representation where this spatial anisotropy has not only a more pronounced character, but is also accompanied by a qualitative change in the *extrinsic* geometry of the time dimension: for, as easily discernible from the comparison of the *left*-hand sides of Figures 2 and 3, the pencil of fundamental conics in the generic case is of the type "a" of Figure 1, whereas that in this particular degenerate case is of the type "b" (as $B_3 \equiv L$ here).

A brief inspection of some other degenerate cases [27,29] reveals further subtleties of this fascinating relation. Thus, the same type of time dimension ("b") is found also in the case when two of the base points coalesce, not on $\mathcal{L}^B$; space now featuring only two dimensions, both being ordinary. The "c" type of time coordinate is encountered if one of the base points lies on $\mathcal{L}^B$ and the other two coincide, off the latter; space again endowed with a couple of dimensions – one ordinary, one extraordinary. The dimensionality of space is further reduced if all the three base points merge, the single spatial coordinate being ordinary or extraordinary and time converted into the type "d" or "e" according as the merger lies off or on the line $\mathcal{L}^B$, respectively. The time dimension can even disappear, as in the case when two of the base points fall on $\mathcal{L}^B$; space, remarkably, retains here its three dimensions, of which two are extraordinary.[3] These examples suffice to see that *a profound connection between the global structure of time dimension and the number, as well as individual character, of spatial coordinates* is an essential element of the structure of our Cremonian pencil-space-times. It is this property that finds its most distinguished and almost ubiquitous manifestations in the already-mentioned realm of the psychopathology of time and space [1–14,21–23]. However, from what we found above it stems that this must also be a feature pertaining to the structure of the physical universe, although in this case its traits are obviously of a much subtler nature, being essentially embodied in a possible delicate non-equivalence among the spatial dimensions themselves (compare the *right*-hand sides of Figures 2 and 3). And although so far successfully evading any experimental, observational evidence, this fact deserves at least serious theoretical attention, especially given recent trends in quests for a "theory of everything" [30]. For in addition to invoking compactified extra dimensions of space to get a sufficiently extended setting for the ultimate unification of all the known types of interaction, we think it is worth revisiting and having a fresh look at (the relation among) their three classical, infinitely-stretched-out relatives, available to our senses.

The final task that remains to be done concerns the *intrinsic* structure of the pencil time dimension. This structure has to conform to our experience of time as consisting of three qualitatively different kinds of event, viz. the past, present and future. Yet, our pencil-time is so far homogeneous, because all proper conics are projectively equivalent. We therefore need to "de-homogenize" the pencil to yield the structure required. In light of the strategy pursued in our planar model [15–23], one of the simplest ways of doing so is to select, in the unprimed projective space, one line, $\mathcal{L}^*$, and attach to it a special status. It is an easy exercise to see that if this line is in a general position, it will cut each of the four "fundamental" planes in a point which does not coincide with any of the base points $B_k$

---

[3] The interested reader may try to extend this analysis to all the remaining degenerate cases, whose complete list and basic group-geometrical description can be found in [29].

($k$=1,2,3) and lies off $\mathcal{L}^B$ as well. Hence, the line $\mathcal{L}^*$ will be incident with a unique line from each of the three "fundamental" pencils in $B_k\mathcal{L}^B$ ($k$=1,2,3) and a unique, in general proper, conic of the "fundamental" pencil in the plane $B_1B_2B_3$ – as depicted in Figure 4. But as for the pencil of fundamental conics, and so time, there is indeed more than meets the eye. For the point at which $\mathcal{L}^*$ meets the plane in question separates the proper conics of the pencil into two disjoint, qualitatively distinct families. One family consists of those conics for which this point is external ("ex-conics"), while the other family comprises the conics having this point in their interior ("in-conics"); the two sets are separated from each other by a *unique* proper conic, the one that incorporates the point (the "on-conic" – the conic drawn bold in Figure 4). This structure is seen to be perfectly compatible with what Nature offers to our senses after we identify the *ex*-conics with the *past* events, the *in*-conics with the events of the *future*, and the unique *on*-conic with the moment of the *present*, the "*now*." It is evident that nothing similar takes place inside a(ny) pencil of lines, because there are no such notions as external and/or internal for a point with respect to a line. So, our spatial pencil-dimensions retain their "homogeneity," as observed.

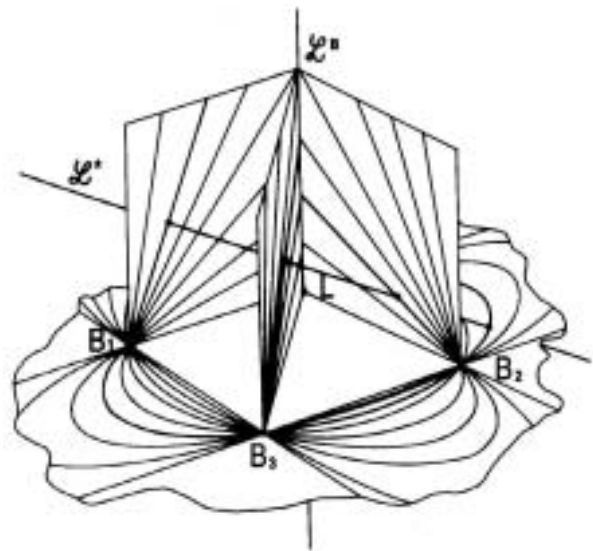

*Figure 4*. A generic line of the projective space, $\mathcal{L}^*$, is incident with three unique, mutually skew, fundamental lines (one from each pencil) and a unique fundamental conic (all the four objects drawn bold).

It is important here to realize that it is only after introducing this standing-out line when the distinction between time and space acquires its desirable, observed form. A natural question emerges: what is the meaning of this special line? We think that the existence of such a line may simply be understood as a possible representation of the observer. It then follows that if there is no observer, there is no "here" and there is no present, past and future either. The converse, however, is not true! That is, there do exist observers that are not (fully) localized in space and whose time dimension exhibits a completely different internal structure. As the attentive reader may already have noticed, this has to correspond to the cases where the line $\mathcal{L}^*$ has a particular position with respect to the points $B_k$ and/or the line $\mathcal{L}^B$, or lies completely in one of the "fundamental" planes. To illustrate the point, let us consider the case where $\mathcal{L}^*$ passes via one of the base points, say $B_1$, being skew with $\mathcal{L}^B$ and not lying in the plane $B_1B_2B_3$. It is clear that in this case every fundamental conic is the on-conic; so, for this particular observer there exists no past/future, *all* the events pertaining solely to the *present*! Moreover, as $B_1$ is the vertex of the pencil of fundamental lines located in the plane $B_1\mathcal{L}^B$, this observer will also find himself/herself to be *in*finitely-stretched-out along the corresponding spatial dimension! Astonishing? Or, rather, weird? Yes, but even more so that this and plenty of even more bizarre, whimsical experiences of space and time are so often found in the narratives of people who find themselves in a profoundly altered state of consciousness and try to share their uncanny experiences with others [1–14,21–23]. To provide the reader with a sense of what such a "strange" space-time experience looks like, we introduce the following fascinating account [31]:

I woke up in a whole different world in which the puzzle of the world was solved extremely easily in the form of a different space. I was amazed at the wonder of this different space and this amazement concealed my judgement, this space is totally distinct from the one we all know. It had different dimensions, everything contained everything else. *I was this space and this space was me*. The outer space was a part of this space, *I was in the outer space and the outer space was in me...*

Anyway, I didn't experience *time, time of the outer space and eons* until the second phase of this dream. In the cosmic flow of time you saw worlds coming to existence, blooming like flowers, actually existing and then disappearing. It was an endless game. *If you looked back into the past, you saw eons, if you looked forward into the future there were eons stretching into the eternity and this eternity was contained in the point of the present.* One was situated in a state of being in which the "will-be" and the "vanishing" were already included, and this "being" was my consciousness. It contained it all...

## 3. Conclusion

Current science is most adept in addressing problems that require technique rather than insight. Yet, when addressing the fundamental issues of the structure of space-time, it is rather insight that matters. The above-outlined theory of Cremonian space-time(s) seems to provide us with both, which is one of its strongest points. It not only offers us a feasible explanation why the Universe features three spatial and one temporal dimension, but also indicates unsuspected intricacies of the coupling between the two. Moreover, it also sheds fresh light on how the physical view of space-time and its experiential counterpart can possibly be interconnected. These properties alone are enough to realize that the theory deserves further serious exploration.

**Acknowledgements**

This work was supported by the NATO Advanced Research Fellowship, distributed and administered by the Fonds National de la Recherche Scientifique, Belgium, and, in part, by the NATO Collaborative Linkage Grant PST.CLG.976850. I would like to thank Mr. P. Bendík for careful drawing of the figures. I am also grateful to my friends Prof. Mark Stuckey (Elizabethtown College) for a careful proofreading of the paper and Dr. Rosolino Buccheri (IFCAI, Palermo) for valuable comments.